# MASS STORAGE MANAGEMENT AND THE GRID

A. Earl[*], P. Clark[†], University of Edinburgh, Edinburgh, Scotland

*Abstract*

The University of Edinburgh has a significant interest in mass storage systems as it is one of the core groups tasked with the roll out of storage software for the UK's particle physics grid, GridPP. We present the results of a development project to provide software interfaces between the SDSC Storage Resource Broker, the EU DataGrid and the Storage Resource Manager. This project was undertaken in association with the eDikt group at the National eScience Centre, the Universities of Bristol and Glasgow, Rutherford Appleton Laboratory and the San Diego Supercomputing Center.

## INTRODUCTION

Management of data and storage resources is a major problem for regional, or Tier 2, computing centres. They are expected to support the computing and storage requirements of scientific research but are hampered by incompatibilities, due to the current lack of standards, between the solutions used by each discipline. The University of Edinburgh has a significant interest in this area, as it houses the storage biased site of the ScotGrid [1, 2] regional centre which supports groups in several disciplines including particle physics and bio-infomatics.

Man power availability, and equipment configuration, allows us to realistically support only one solution; however, we believe that this is not an intrinsic limitation, but a result of the limited development effort which has been spent on interfacing the various existing solutions. In this paper we outline two systems which are popular with our user communities and our progress in ensuring that they are interoperable.

*The Storage Resource Broker*

The Storage Resource Broker (SRB) [3] is a software suite developed at the San Diego Supercomputing Center (SDSC) to provide access to distributed data storage facilities using a centralised metadata management facility. SRB uses *media drivers* to provide an interface between it and various databases, storage management systems and file systems. These media drivers allow the rapid adoption of new resource types and promote transfer protocol and data type negotiation between client applications and storage.

The central metadata catalogue (MCAT) has been developed for use with relational databases including Oracle, IBM DB2 and PostgreSQL. Performance and reliability can be improved by using the features of these industrial strength database systems. The centralised nature of this system allows for data mining and efficient system administration. Work on SRB Zones has improved the systems ability to communicate between different projects with different MCATs.

*The Storage Resource Manager*

The Storage Resource Manager (SRM) is a set of specifications for providing a Grid interface to storage management systems of various types including disk, tape, and hybrid. There is a specification for Unix based systems and a further one based on Web Services. These specifications have been developed by collaborators in Europe and the United States as part of a Global Grid Forum working group and implemented on various platforms. For the purposes of this paper we refer to the version [4] provided by the Lawrence Berkeley National Laboratory (LBNL) which provides both a disk (DSM) and a tape (TSM) implementation. Both of these are used as part of their hierarchical storage manager (HSM) system.

The SRM specification does not detail how metadata about the files and records being stored in the system should be managed. This is left to other $3^{rd}$ party services on the assumption that different groups will wish to take their own situation into account when designing the information service.

*The GMCat project*

The GMCat software [6] was developed at the University of Bristol and used in the CMS 2004 Data Challenge [8] to provide a consistent mapping of files stored on the LHC Computing Grid (LCG) and SRB. The current version is based on Web Services and replicates the LRC interface, providing a link between SRB (datanames) and LCG (SURLs) namespaces. This allows files catalogued in the SRB MCat to appear in the Replica Location Service (RLS) hierarchy, and therefore become visible to LCG.

[*] aearl@ed.ac.uk
[†] p.j.clark@ed.ac.uk

# THE SRM2SRB PROJECT

While the GMCat project has made significant progress toward providing interoperability between SRB and the RLS, it does not address the problem of accessing the stored data, other than defining SRB as an access protocol when data is registered with the RLS. Although it is possible to use generic Grid software, such as Globus, to access this data if it is read-only and maintained in an area accessible by the user accounts that the Grid gatekeeper maps requests to, in more complex systems, where there are pooled accounts and data must be modified, the problems of scalability and security soon become apparent. The aim of the SRM2SRB project is therefore, to identify these issues and provide solutions.

## Existing Systems

Figure 1 shows a simple SRB system in which there exist: SRB Servers, which are Unix daemons resident on any resource which provides storage capabilities; the Metadata Catalog (MCat) which is a relational database; and the SRB Master, which provides the functionality of an SRB Server with the added ability of being able to communicate with the MCat. When users wish to add, retrieve or manipulate files in the system they communicate with the SRB Server on their local resource which then queries the MCAT through the SRB Master.

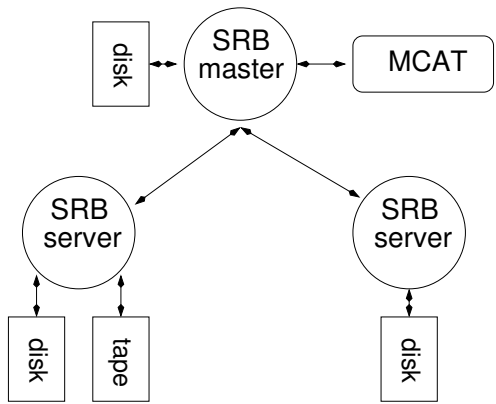

Figure 1: A simple SRB system

At each of these resources there is also an SRB Vault. This is an SRB managed directory in which replicas of the files in the system are stored to improve performance and reliability. A trivial method of accessing these files would be to map users onto the *srbadmin* account however this would break the file permissions stored in SRB. Fortunately SRB supports Grid Security Infrastructure (GSI) authentication through the use of Grid Certificates which are not reliant on the user account the request comes from. This involves more work for the system administration during the initial configuration of the system but is likely to have significant benefits for the users who will not require a specific SRB account.

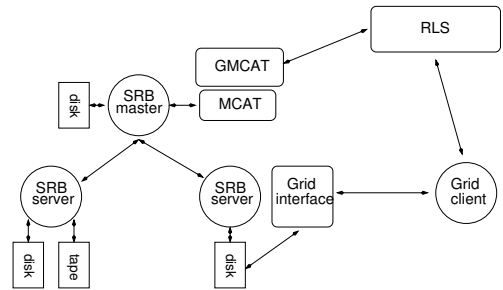

Figure 2: The existing SRB / SRM situation

The LBNL SRM implementation provides the GridFTP functionality which SRB currently lacks, although there has been some work conducted in this area by SDSC. This is discussed in more detail in the next section.

## Designing SRM2SRB

Several potential options as to how we could achieve an SRM Interface to SRB present themselves and within these we can clearly see a progression between providing a minimal service and developing an ideal solution. We could create a completely new SRM interface, we could adapt the existing LBNL TSM software, or we could create a new driver specifically for SRB.

With the first option, creating a totally new SRM is beyond our current man-power capabilities and does not have enough support, or immediate need, to make an attractive proposal. SRM functionality, such as file pinning and space reservation, are not supported by SRB and it would, therefore, not achieve complete SRM compatibility. The benefits of code reuse would not be felt from developing a new SRM implementation while the maintenance and support issues would increase the demands on our own system administrators.

A more realistic option, and one which we believe could be an interesting proof of concept project, would be to adapt the existing tape driver to fetch files from SRB, which are then transferred to the user using the disk driver. This option has obvious performance issues as it would require us to copy data to disk irrespective of whether it exists there already or not. This is demonstrated in Figure 3. Further consideration of this option however suggests that there may be benefits.

We are currently assuming that not all sites which are using SRB wish to migrate to SRM for whatever reason. Therefore, although GMCat will have registered all systems and their contents with RLS, not all of them will be accessible to users. At this point can we also identify the problem of a potential mismatch between MCat and RLS. The MCat maintains information about file access

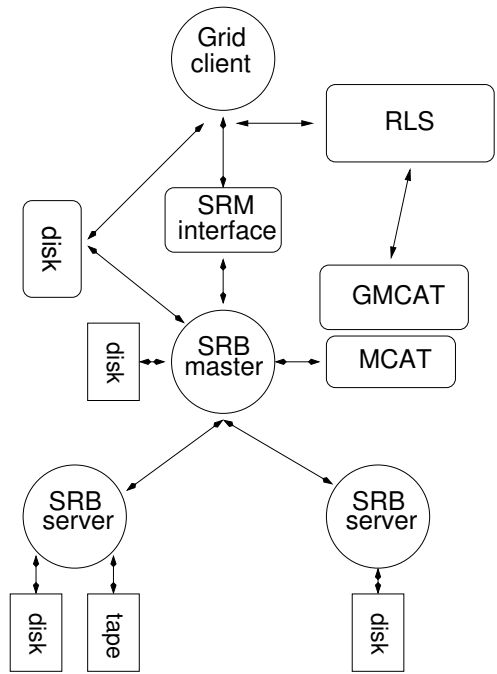

Figure 3: The SRM2SRB design

permissions such as which users can read, write and delete files while RLS does not.

Because of this, users can locate a file using the RLS and only discover that they are unable to retrieve it when they request it assuming that they are using the SRB drivers. In situations such as the trivial case, outlined in the previous section, where Grid software directly retrieves files rather than using the SRB API, the potential for an unstable system is significant.

With this solution, having a central site at which all requests can be processed would normally suggest a bottleneck, but our experience has shown that the central SRB site usually has significant storage, processing and networking capabilities and maintains replicas of a large proportion of the data. Assuming that there are only a small number of sites which require the data and are not running SRB, this is a manageable situation.

With the third option we could make use of the modular design of the LBNL SRM implementation, where there is a separate SRM interface and driver which communicate using CORBA, to create a new driver specifically to support SRB. With both this, and the previous option, issues such as file permissions and modification notifications will be handled by SRB, as we are using its API, and will be discovered by RLS when queried. With this scenario the driver would have the advantage that data would not necessarily need to be staged to tape assuming that protocols other than GridFTP can be used for data transfer, and the full SRB functionality, such as Data Cutter could be utilised to improve performance.

## CONCLUSIONS

From the work we have conducted so far it seems obvious that treating SRB as a tape system is not a viable long term option, as the performance issues and management complications greatly outweigh the potential benefits of code reuse. This is balanced by the fact that developing an entirely new SRM system to support SRB has issues, as the two systems have fundamental differences in what they are trying to achieve and the systems they are intended to be used for.

Our current road map for development at Edinburgh is to have produced a working, and tested, version of a driver compatible with the LBNL SRM interface based on the tape driver by the end of November. If this proves useful further work to produce an SRB specific driver will be considered. Details of this are available at http://www.ph.ed.ac.uk/ aearl/srm2srb/.

## ACKNOWLEDGEMENTS

We would like to thank the Particle Physics and Astronomy Research Council (PPARC) for studentship PPA/S/E/2001/03338. Thanks also to Wayne Schroeder of SDSC for his assistance with SRB and the National eScience Centre visitor programme for making Wayne's visit possible.

Arun Swaran Jagatheesan, Owen Maroney, Tim Barrass, and Simon Metson deserve special mention for their work and assistance with the designs used in this paper and their respective work on SRB, SRM and GMCat also Owen Synge and Jens Jensen for their work on the EU DataGrid Storage Element. Many thanks to Simon for his assistance with this paper and his knowledge of the metadata systems involved.